\documentclass{aa}
\usepackage{txfonts}
\usepackage{natbib}
\usepackage{graphicx}

\bibpunct{(}{)}{;}{a}{}{,}

\begin{document}

\defcitealias{gir04a}{G04}

\titlerunning{The jet of Mrk 501 from $10^6 R_S$ down to a few $10^2$.  }
\authorrunning{Giroletti et al.}

\title{The jet of Markarian 501 from millions of Schwarzschild radii down to a few hundreds. } 

\author{M. Giroletti\inst{1}
  \and  G. Giovannini\inst{1,2}
  \and	W.D. Cotton\inst{3}
  \and	G.B. Taylor\inst{4}
  \and	M.A. P\'erez-Torres\inst{5}
  \and	M. Chiaberge\inst{1,6}
  \and	P.G. Edwards\inst{7}
}

\institute{INAF Istituto di Radioastronomia, via Gobetti 101, 40129, Bologna, Italy
\and Dipartimento di Astronomia, Universit\`a di Bologna, via Ranzani 1, 40127 Bologna, Italy
\and National Radio Astronomy Observatory, 520 Edgemont Road, Charlottesville, VA 22903-2475, USA
\and Department of Physics and Astronomy, University of New Mexico, 800 Yale Boulevard NE, Albuquerque, NM 87131, USA
\and Instituto de Astrof\'{\i}sica de Andalucia, IAA-CSIC, Apdo. 3004, 18080 Granada, Spain
\and Space Telescope Science Institute, 3700 San Martin Drive, Baltimore, MD 21218, USA
\and Paul Wild Observatory, CSIRO Australia Telescope National Facility, Narrabri, NSW 2390, Australia}

\date{Received / Accepted }

\abstract {} {The TeV BL Lac object Markarian 501 is a complex, core dominated
radio source, with a one sided, twisting jet on parsec scales. In the
present work, we attempt to extend our understanding of the source physics to
regions of the radio jet which have not been accessed before. }  {We present
new observations of Mrk 501 at 1.4 and 86 GHz. The 1.4 GHz data were
obtained using the Very Large Array (VLA) and High Sensitivity Array (HSA) in
November 2004, in full polarization, with a final r.m.s.\ noise of $25\,
\mu$Jy/beam in the HSA total intensity image; the 86 GHz observations were
performed in October 2005 with the Global Millimeter VLBI Array (GMVA),
providing an angular resolution as good as $110\, \mu\mathrm{as} \times \, 40
\, \mu\mathrm{as}$.}  {The sensitivity and resolution provided by the HSA make
it possible to detect the jet up to $\sim$700 milliarcseconds (corresponding to
a projected linear size of $\sim$500 pc) from its base, while the superior
resolution of the 86 GHz GMVA observations probes the innermost regions of the
jet down to $\sim$ 200 Schwarzschild radii. The brightness temperature at the
jet base is in excess of $6 \times 10^{10}$ K.  We find evidence of limb
brightening on physical scales from $\la1$ pc to $\sim 40$ pc. Polarization
images and fits to the trend of jet width and brightness vs.\ distance from the
core reveal a magnetic field parallel to the jet axis.}  {}


\maketitle

\section{INTRODUCTION}

The study of extragalactic radio jets is an important area in astrophysics. In
radio loud sources, jets contribute a large fraction of the total radiated
power, and sustain the formation of energetic kiloparsec scale lobes.  While
observational properties of jets are widely differentiated, they are present in
high and low power sources, with some common features; on the parsec scale,
they are relativistic in both types and they are also intrinsically identical
in beamed and misaligned sources.  \citet{gio01} have shown that the Lorentz
factor in the parsec scale jet of low power FRI radio galaxies as well as of
more powerful FR IIs are both in the range $\Gamma = 3 - 10$. With these
values, jets also appear intrinsically identical in beamed (BL Lac objects) and
misaligned sources (FRI radio galaxies), if the former have jet axis oriented
at an average viewing angle of $\langle \theta \rangle = 18^\circ \pm 5^\circ$
\citep{gir04b}.

In the present paper we focus on the jet structure of the BL Lac source
Markarian 501. This object is highly active and well-studied at all
frequencies. Its activity and variability at high energy \citep[as high as TeV
regime,][]{qui96} seems to require high Doppler factors and consequently a
small angle to the line of sight. In the radio band, centimeter VLBI
observations have revealed a clear limb-brightened structure, beginning in the
very inner jet, suggestive of a dual velocity structure \citep[~hereinafter
G04]{gir04a}. The complex limb-brightened structure makes component
identification problematic and multi-epoch attempts to measure pattern speed
conclude that it is not well defined \citepalias{gir04a} or in any case at most
subluminal \citep{edw02}. These seem to be common features in TeV blazars
\citep{pin04,gir06}, and theoretical models have been proposed to reconcile
them with the very high energy emission \citep{ghi05,wan04}.

However, the results obtained so far still leave some major questions
unanswered. For example, it is not at all clear whether the jet velocity
structure is intrinsic or produced by the interaction with the surrounding
medium. We want to understand why the properties of the radio jet on parsec
scales are different from those needed to explain the $\gamma-$ray emission in
Mrk~501, as well as in other TeV blazars. Moreover, a change in regime must
occur on much larger scales, since the large scale structure of the source is
known to be symmetric rather than one-sided \citep[e.g.,][]{ulv83}. We want to
investigate if this transition is smooth and what is the configuration of the
magnetic field in the outer jet, which needs sensitive images in total
intensity and polarization.

In order to search for an answer to such questions, we need to go beyond the
capability of the instruments available for ordinary centimeter wavelength
VLBI, which provides information only for the region between $\sim 1$ and $\sim
100$ milliarcseconds. Smaller and larger scale regions remain inaccessible
because of inadequate resolution and sensitivity, respectively.

Improvements in the technical and organizational issues are now offering to
astronomers VLBI arrays of unprecedented resolution and sensitivity, such as
the High Sensitivity Array (HSA\footnote{{\tt http://www.nrao.edu/hsa/}}), and
the Global mm-VLBI Array (GMVA\footnote{{\tt
http://www.mpifr-bonn.mpg.de/div/vlbi/globalmm/}}).  Thanks to its proximity
and brightness, Mrk 501 is an ideal laboratory for experiments using these
advanced VLBI techniques: it is at $z = 0.034$ (1 mas = 0.67 pc, using $H_0 =
70$ km s$^{-1}$ Mpc$^{-1}$); the total flux density at 5\,GHz is $S_5 =
1.4$\,Jy; the Schwarzschild radius for its central black hole is estimated
around $1 R_S = 10^{-4}$\,pc ($1.4 \times 10^{-4}$\,mas), if we adopt
$M_\mathrm{BH} = 10^9\,M_\odot$ \citep{rie03}.  Using these new facilities, we
can therefore access regions never studied previously: the jet base with the
GMVA, and the faint, resolved jet region at $>100$\,mas with HSA.

\begin{figure}
\resizebox{\hsize}{!}{\includegraphics{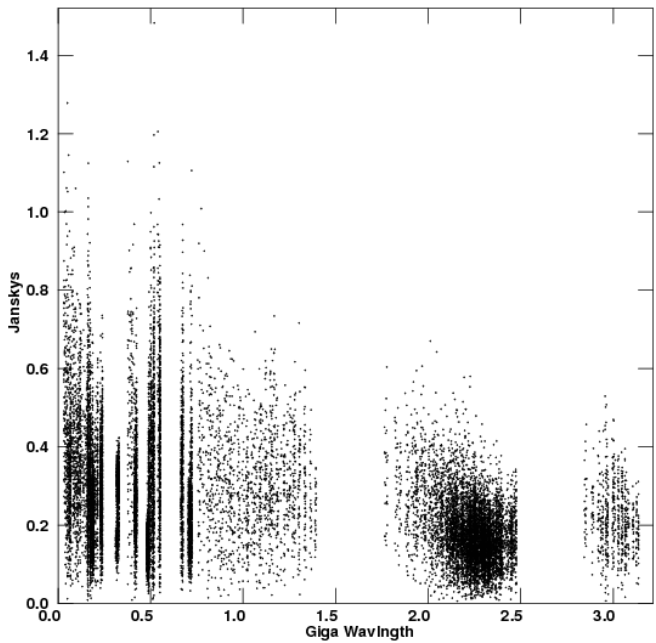}}
\caption{Visibility amplitude vs.\ $(u,v)-$radius for the GMVA
observations.\label{f.uv.gmva}}
\end{figure}

In \S \ref{sec:observations} we describe the instruments used for our new
observations, along with the data reduction methods required by high
frequency. Results are presented in \S 3 and discussed in \S
\ref{sec:discussion}. We present our conclusions in \S \ref{sec:conclusions}.

\section{OBSERVATIONS}
\label{sec:observations}

\subsection{High Sensitivity Array observations}

We observed Mrk 501 with the HSA at 1.4\,GHz on 26 Nov 2004. The HSA is
obtained by combining in the same array the 10 VLBA antennas and other
sensitive elements, i.e. the Green Bank Telescope (GBT, 100 m.), the phased VLA
($27 \times 25$ m.), Arecibo (300 m.), and Effelsberg (100 m.). Even without
Arecibo, whose declination limits do not allow it to observe Mrk~501, the
collecting area is increased by a factor of 7 over the VLBA alone.  The
sensitivity was also improved thanks to a large recording rate (256\,Mbps) and
a long integration time (8\,hrs). The Effelsberg telescope was in the
experiment for the first 5 hrs; some failures affected SC, FD, MK, and the VLA
during part of the observation.

We reduced the data in the standard way in AIPS, using 3C345 as a fringe
finder, OQ208 as a leakage calibrator, and 3C286 for the EVPA
calibration. Final images were produced both in AIPS and Difmap, with different
weighting schemes. The source structure is complex, with a strong peak ($\sim
0.7$\,Jy) and significant diffuse emission ($\sim 0.8$\,Jy). Although this
prevents us from reaching the thermal noise, we still achieve in our best image
a dynamic range as good as 30,000:1, with a noise level of $\sim 25\,\mu$Jy
beam$^{-1}$ ($1\sigma$).

During the HSA observations, the VLA (used as phased array) was in the A
configuration.  We obtained the internal VLA data and calibrated and reduced in
the standard way to obtain also a VLA image of Mrk~501.

\begin{figure} 
\resizebox{\hsize}{!}{\includegraphics{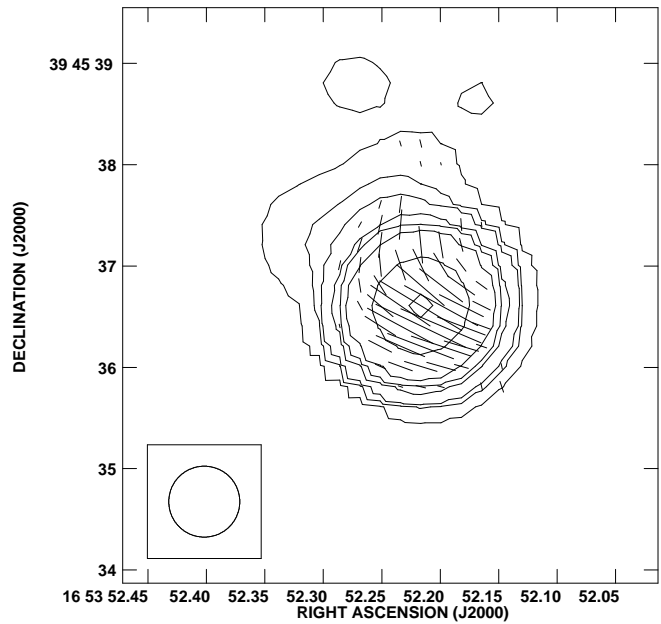}}
\caption{VLA image of Mrk 501. The peak is 1.7 Jy beam$^{-1}$, and the contours
are traced at $(1, 5, 10, \dots) \times 1.0$ mJy beam$^{-1}$. The restoring
beam is circular with a FWHM of $0.7\arcsec$. Sticks represent polarization
vectors, with a scale of 1\arcsec = 8.3 mJy beam$^{-1}$.\label{f.vla}}
\end{figure}

\subsection{Global mm-VLBI observations}

\begin{figure*}
\sidecaption
  \includegraphics[width=12cm]{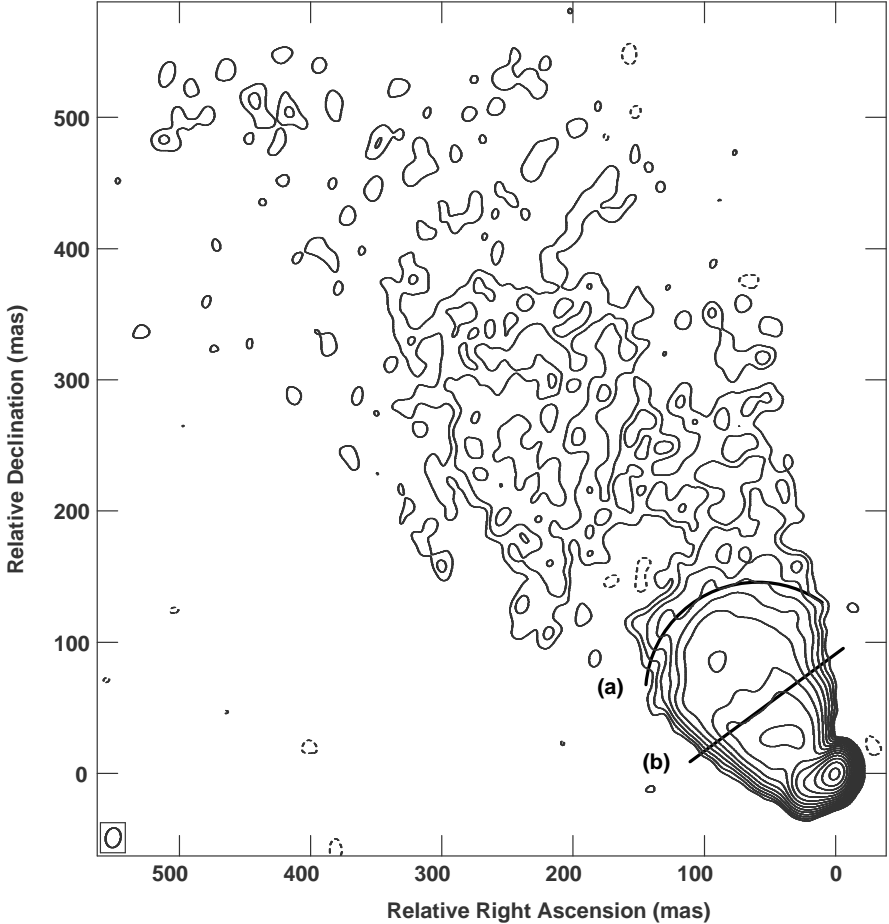} 
\caption{High Sensitivity Array image of Mrk 501. The peak is 762 mJy
beam$^{-1}$, and the contours are traced at $(-1, 1, 2, 4, \dots) \times 0.075$
mJy beam$^{-1}$ ($3\sigma$ noise level). As a result of natural weighting with
a Gaussian taper at 18 M$\lambda$, the restoring beam is $15.1 \times 11.4$
(mas $\times$ mas) in PA $-11.3^\circ$. We mark with letters the region of
sharp brightness decrease $(a)$ and of limb brightening $(b)$ in the extended
jet. See Fig.~\ref{f.lb} for the brightness profile across the
jet.\label{f.largest}}
\end{figure*}

Millimeter VLBI permits a much higher angular resolution than ground or space
based VLBI at centimeter wavelengths. Moreover, it offers the possibility to
study emission regions which appear self-absorbed at longer wavelengths, with
important consequences for our understanding of the physical processes in AGNs
in the vicinity of supermassive black holes. After years of continuous
development and technical improvement, the GMVA is now able to provide good
quality images in the 3mm band, with an angular resolution of a few tens of
micro-arcseconds \citep{kri06a}.

We observed Mrk 501 on 14 Oct 2005 with the Global mm-VLBI Array. This
experiment tested the sensitivity limits of the array, since on the basis of
the observed centimeter wavelength flux density and spectral index
\citepalias{gir04a}, Mrk 501 was expected to be only a few hundred mJy at this
frequency.

The standard frequency was 86.198 GHz, with 16 IFs of 8 MHz bandwidth each, 2
bit sampling in left circular polarisation (LCP). The participating telescopes
were Effelsberg, Pico Veleta, the Plateau de Bure interferometer, Onsala,
Mets\"ahovi, and 8 VLBA stations (i.e.\ all except Saint Croix and
Hancock). The European telescopes observed for $\sim 9$ hours and the American
ones joined in for the last $\sim 6$ hours (Mauna Kea only for the last $\sim
4$ hours); the telescopes at Mets\"ahovi and North Liberty failed.

The calibrator 3C345 was readily detected with good signal-to-noise ratio. From
the fringe fitting of 3C345 we determined rates and single-band delays, and
applied them to the whole data set. We obtained an image of 3C345 and found it
to be in agreement with published images of comparable or slightly lower
resolution \citep{lob00,lis05}.  At this stage, it was then possible to fringe
fit Mrk 501 itself, averaging over the IFs, using a solution interval as long
as the scan, and setting a SNR threshold of 3.0. Mrk~501 was well detected not
only between large European apertures but also on baselines to the smaller VLBA
antennas. Solutions that were obviously bad were edited out using SNEDT, and
the data were subsequently frequency averaged.  Final self-calibration and
imaging were done in Obit \citep{cot08}.

The final amplitude vs.\ $(u,v)-$distance plot is shown in
Fig.~\ref{f.uv.gmva}. The coverage is good in the short baseline range and much
sparser in the outer part of the $(u,v)-$plane. Due to the failure of the
easternmost VLBA antenna, there is also a large gap in between the short and
long baseline domains. A large baseline noise is visible; however, significant
emission in the short baselines is clearly present. An image of the calibrator
3C345 and a spectral plot of the resulting phases vs.\ spectral channels for
visibilities of Mrk 501 are shown in \citet{gag06}.

\section{RESULTS}
\label{sec:results}

\subsection{The kpc scale structure}

On kiloparsec scales, Mrk 501 is core dominated with a two sided extended
structure visible as well, extending in PA $\sim 45^\circ$ for more than
30\arcsec\ on both sides of the core \citep{ulv83,kol92,cas99}. It is
straightforward to identify this structure with the symmetric extended emission
characteristic of a radio galaxy and to infer an orientation near to the line
of sight, in agreement with what is expected from a BL Lac source. However, the
symmetric emission implies that at this distance from the core no relativistic
jet remains.

Thanks to the VLA data available as a byproduct of the HSA observations, we
obtained a higher resolution VLA image of Mrk~501 (see Fig.~\ref{f.vla}).  The
phased array image is dynamic range limited ($\sim 10000:1$), and it shows a
one-sided emission with a short jet like structure in the same PA as the
extended symmetric structure.  From this one-sided emission we can derive
constraints on the jet velocity ($\beta c$) and orientation ($\theta$) with
respect to the line of sight. At $2\arcsec$ we have $\beta \cos \theta > 0.36$
and at $1\arcsec$, $\beta \cos \theta> 0.63$. This result implies that at 0.67
kpc (projected) from the core the jet is still at least mildly relativistic
($\beta > 0.63$).

\subsection{The extended jet}
\label{s.jet}

We obtain a detailed look at the jet of Mrk~501 from the deep VLBI observations
with the HSA. We show in Fig.~\ref{f.largest} a tapered image, where baselines
longer than 18 M$\lambda$ have been significantly down-weighted to increase the
signal to noise ratio of the low-surface brightness emission. We achieve a
$1\sigma$ r.m.s.\ of $\sim 25\, \mu$Jy beam$^{-1}$ and emission is revealed on
the main jet side up to a distance of $\sim 700$ mas from the core, i.e., five
times further than detected in any previous VLBI observation. No emission is
detected on the counter-jet side at the level of $3\sigma$ noise. In Table
\ref{t.jcj}, we give at some selected distances (Col.\ 1) the jet brightness
(Col.\ 2) and the corresponding lower limits to the jet/counter-jet brightness
ratio (Col.\ 3). The minimum required velocity ($\beta_{\rm min}$) and largest
allowed viewing angles ($\theta_{\rm max}$) are then reported in Cols.\ 4 and
5, respectively. The jet opening angle remains constant ($\phi_\mathrm{j} \sim
40^\circ$) after the well known bend at 30 mas.

\begin{figure}
\resizebox{\hsize}{!}{\includegraphics{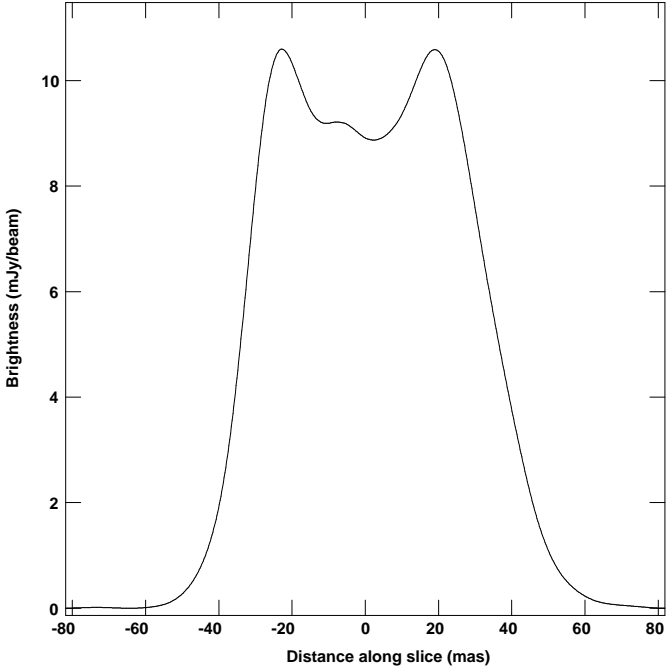}}
\caption{Jet brightness profile across the slice shown in Fig.~\ref{f.largest}
(center is at RA 16$^h$ 53$^m$ 52$^s$.2435035, Dec.\ +39$^\circ$ 45\arcmin
36\arcsec.906318, PA $= -53.4^\circ$).\label{f.lb}}
\end{figure}

\begin{table}
\caption{Jet/counter-jet brightness ratio}
\label{t.jcj}
\centering
\begin{tabular}{rcccrl}
\hline\hline
 $r$   &  $B_J$ &  &  &  $\theta_{\rm max}$ &   \\
 (mas) & (mJy beam$^{-1}$) & $R_{\rm min}$ & $\beta_{\rm min}$ & ($^\circ$) & Notes \\
 (1) &  (2) &  (3) &  (4) &  (5) &  (6) \\
\hline
12  & 180  & 7200 & 0.94 & 19 & Inner jet \\
21  & 105  & 4200 & 0.93 & 21 & Jet bend \\
62  & 21.4 & 856  & 0.87 & 29 & \\
128 & 3.5  & 140  & 0.76 & 41 \\
284 & 0.26 & 10.4 & 0.44 & 64 \\ 
464 & 0.20 & 8.0  & 0.39 & 67 & \\
706 & 0.12 & 5.0  & 0.31 & 72 & confused with noise \\
\hline
\end{tabular}
\end{table}

The jet brightness shows a decrease with increasing jet distance from the core;
the space distribution is quite uniform, i.e., no prominent knot is present in
the extended jet. The most noteworthy features in Fig.~\ref{f.largest} are the
relatively sharp brightness decrease at $\sim 100$\,mas (marked with $a$),
which could correspond to a shock region, and the jet limb brightening across
the slice at $\sim 60$\,mas from the core (marked by $b$). The jet brightness
profile across this slice is shown in Fig.~\ref{f.lb} and, although less
conspicuous, is similar to the structure visible in the inner jet in higher
resolution images \citepalias{gir04a}.

Besides these features, the jet presents an uniformly distributed flux
density. If we consider only the region above the $3\sigma$ noise level, 73.2\%
of the pixels in the image have a brightness between 75 and 300 $\mu$Jy
beam$^{-1}$. Significant peaks are not present and local maxima can be related
to small increases in the jet emissivity but also to artifacts brought about by
the image reconstruction process. This seems to be a characteristic of the jet
of Mrk 501 on all scales; even in the inner jet, images with high resolution
\citepalias{gir04a} tend to show a uniform brightness distribution rather than
the compact knots observed in other AGN jets.

In \citetalias{gir04a}, we modeled the jet intensity of Mrk~501 as a function
of jet velocity and radius, using the formulas for an adiabatically expanding
jet derived by \citet{bau97} in the case of relativistic motion. The most
sensitive observations available in \citetalias{gir04a} allowed us to study the
jet only to a distance of $\sim 100$ mas, still allowing for a degeneracy
between magnetic field orientation and jet velocity. Thanks to the HSA data, we
can now extend this argument to a distance of almost 500\,mas.

\begin{figure}
\resizebox{\hsize}{!}{\includegraphics{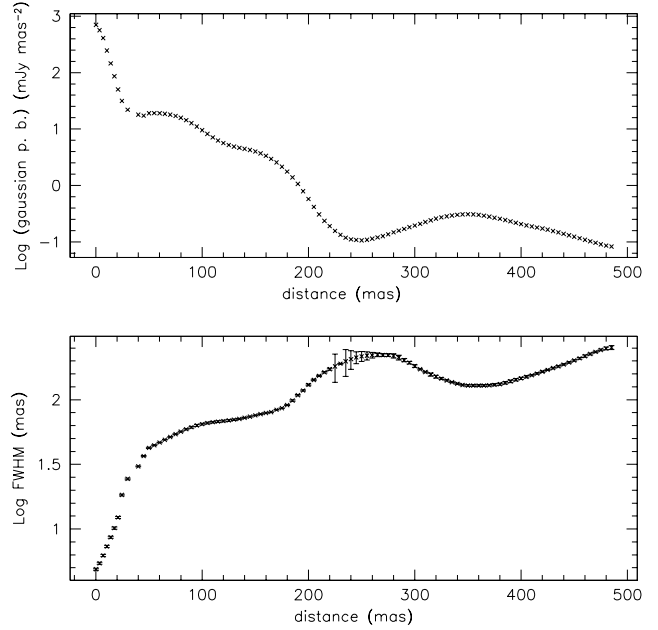}}
\caption{Jet peak brightness and FWHM vs.\ distance from the core obtained from
Gaussian fits. A boxcar filter (50 mas) has been applied to smooth the data at
$r>50$ mas.\label{f.gaussians}}
\end{figure}

We derived brightness profiles across the jet using the AIPS task
\texttt{SLICE} on the tapered HSA image for the extended jet, obtaining one
slice every 5 mas in PA$=-56^\circ$. Using the AIPS task {\tt SLFIT}, we fitted
single Gaussian components to each profile. We show the resulting data as
functions of the distance from the core in Fig.~\ref{f.gaussians}.  The fit
could be done unambiguously in most cases, although some slices presented some
deviation from a pure single Gaussian profile. However, the difference between
the area subtended by the profile and the fit is generally smaller than 5\% and
only a couple of fits (at $\sim 200$ mas from the core, marked by the larger
error bars) had to be rejected. At a distance larger than $\sim 450$\,mas from
the core, the best fit FWHM starts decreasing; we ascribe this behavior to the
insufficient brightness at the jet edges, and do not consider any slice at
$r>480$ mas in our analysis. Moreover, we note that for the same reason -- and
because of the limb brightening of the jet -- even at smaller distances the
actual jet FWHM is in some cases larger than the best fit one. The implication
of this effect are discussed below. In the inner jet ($r<30$ mas), we keep the
fit from \citetalias{gir04a}, whose better resolution is important in this
region where the jet is not transversely resolved in the present image.  The
best-fit peak brightness and jet FWHM have then been smoothed with a boxcar
filter (50 mas wide, for points at $r>50$ mas) to suppress local noise.

\subsection{The core and inner jet structure}

\begin{figure} \center
\resizebox{\hsize}{!}{\includegraphics{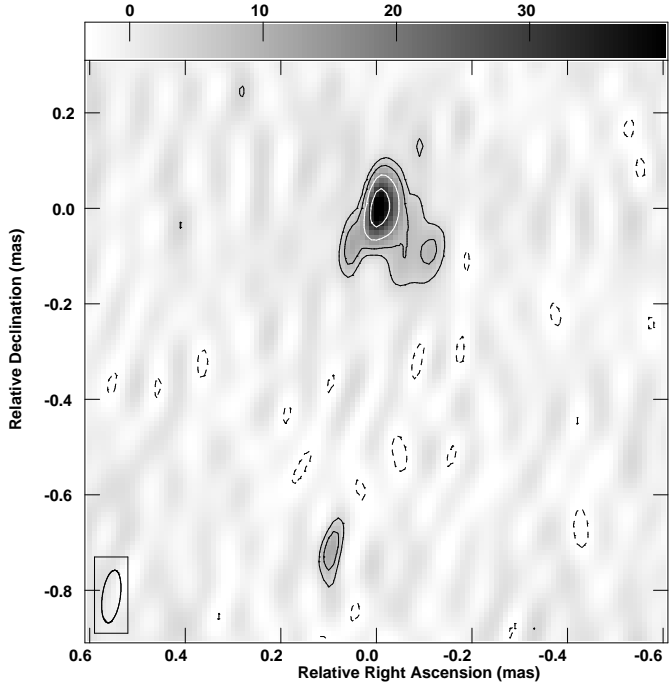}}
\caption{Mrk 501 at 86 GHz; the restoring beam is $110\, \mu$as $\times 40\,
\mu$as in PA $-8^\circ$. The peak is 45 mJy beam$^{-1}$, and the contours are
traced at $(-1, 1, 2, 4, 8) \times 4.0$ mJy beam$^{-1}$. The $1\sigma$ noise
level is $\sim 1.5$ mJy beam$^{-1}$. The grey scale flux range is $-3.0$ to 40
mJy beam$^{-1}$.\label{f.gmva}}
\end{figure}

In Fig.~\ref{f.gmva}, we show our Global mm-VLBI Array image of Mrk 501 at a
resolution of 110 $\mu$as $\times \, 40 \, \mu$as (beam FWHM, PA
$-8^\circ$). Mrk~501 is clearly detected at 3 mm and it is dominated by a
compact, prominent component, $\sim 45$ mJy beam$^{-1}$ peak brightness. The
visibility data suggest that there is a fair amount of extended emission,
although the $(u,v)$ coverage is not ideal and it is extremely difficult to
image it.

\begin{figure} \center
\resizebox{\hsize}{!}{\includegraphics{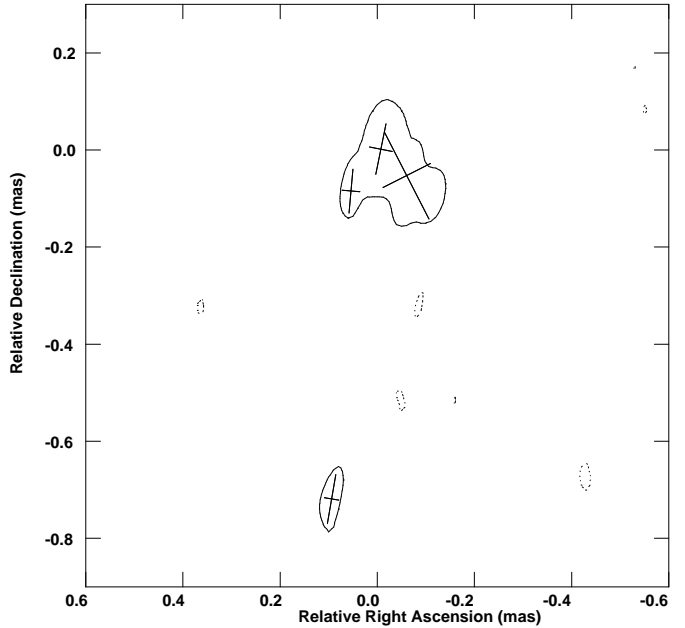}}
\caption{Results from the model-fit with Gaussian components to the image
plane, overlayed with contours traced at ($-4.5, 4.5)$ mJy beam$^-1$. The
crosses mark the position, major and minor axis, and position angle of each
component.\label{f.mf}}
\end{figure}

In our clean image and with modelfitting, we recover $\sim 110$\,mJy in the
core region, including a jet-like feature in PA $144^\circ$ and some more
diffuse emission in PA $\sim -135^\circ$. A tentative jet knot ($\sim 7\sigma$)
is also visible 0.73\,mas south of the core (PA $172^\circ$). The features in
the image plane can be described by model-fitting with four Gaussian
components. These components are shown overlayed to the $3\sigma$ total
intensity lowest contour in Fig.~\ref{f.mf}; each component is represented by a
cross with major and minor axis equal to its FWHM, with the major axis aligned
along the component's position angle. Quantitative results from model fitting
are reported in Table~\ref{t.model}, where $r$ and $\theta$ are the polar
coordinates of the component (re-referenced to the core position),
$b_\mathrm{maj}$, $b_\mathrm{min}$, and $b_\phi$ are the deconvolved major and
minor axis of the component and its position angle, $P$ and $I$ the peak
brightness and the total flux density.

Visibility model-fitting in Difmap provides a reduced $\chi^2 =1.14$ with this
model. The only quantity that has a significant uncertainty in the best-fit
(around 10\%) is the total flux density $I$, while nominal errors on the
component positions are typically much less than 10 $\mu$as.

Since such values are unrealistically small, we have estimated independent
uncertainties on $r$ taking into account two basic parameters for each
component: (1) the peak flux density $P$ and (2) its compactness. In simple
words, the uncertainty brought about by noise in the visibility data (i.e.\
scatter of the $(u,v)-$points) will affect faint diffuse components much more
than bright compact ones. Therefore, we estimate the uncertainty on the
position of each component using the following formula:

$$\Delta r = \frac{1}{2} \frac{\sqrt{b_\mathrm{maj} \times
b_\mathrm{min}}}{P/3\sigma}$$

where $\sigma$ is the image local noise; the formula is therefore related to
the SNR of the component in such a way that the position of a $3\sigma$ feature
is not known to better than its mean angular radius.  The uncertainties
reported in Col.\ (2) are calculated in this way and they have been added in
quadrature to that on the core position (0.004 mas), which is taken as a
reference.

\begin{table*}
\caption{Deconvolution of Gaussian component fit to the 86 GHz
image.\label{t.model}}
\centering
\begin{tabular}{rrrrrrr}
\hline\hline
 $r$ & $\theta$ & $b_\mathrm{maj}$ & $b_\mathrm{min}$ & $b_\phi$ & $P$ & $I$ \\
 (mas) & ($^\circ$) & (mas) & (mas) & ($^\circ$) & (mJy         & (mJy) \\
       &            &       &       &            & beam$^{-1}$) & \\

 (1) &  (2) &  (3) &  (4) &  (5) &  (6) &  (7) \\
\hline
  0.00 &      0.0 &   0.032  & $<0.048$ &  169 & 40.2  & $48.7 \pm 1.5$ \\
 $ 0.08 \pm 0.03$ & $-$135.6 &   0.181  &   0.078  &   27 &  9.7  & $51.0 \pm 4.3$ \\
 $ 0.11 \pm 0.01$ &    144.4 & $<0.090$ & $<0.036$ &  175 & 11.4  & $ 9.0 \pm 1.1$ \\
 $ 0.73 \pm 0.01$ &    172.0 & $<0.102$ & $<0.030$ &  170 & 12.9  & $ 9.5 \pm 1.1$ \\
\hline
\end{tabular}
\end{table*}

The brightest component, which we identify with the core visible at centimeter
wavelengths, is still unresolved at 86 GHz. We then use our deconvolved size of
this component to give an upper limit to the dimension of the jet base, and a
lower limit to its brightness temperature. At $z=0.034$, 1\,mas\,=\,0.67\,pc,
therefore the deconvolved angular size of the GMVA core corresponds to 0.021
$\times$ 0.032\,pc. The black hole mass for Mrk~501 is estimated around
$M_\mathrm{BH} = 10^9\,M_\odot$ \citep{rie03}, which implies a Schwarzschild
radius $R_S = 1.0 \times 10^{-4}$ parsecs. This means that the radio emission
originates in a region that is smaller than 210 $\times$ 320 R$_S$.

We derive the brightness temperature of this region from the following formula:

$$T_B = \frac{B}{2k} \lambda^2$$

In our observations $\lambda=3.5$\,mm; moreover, to derive $B$ in MKS units, we
calculate that 1 beam $= 7.52\times10^{-17} ab$ ster, where $a$ and $b$ are the
major and minor semi-axis of the deconvolved component, in mas. Therefore:

$$T_B = \frac{1.32 \times 10^{-13} (a b)^{-1} B_\mathrm{mJy}}{2 \times 1.38
\times 10^{-23}} \times 12.1 \times 10^{-6} K$$

i.e.,

$$T_B = 5.8 \times 10^4 \times B_\mathrm{mJy} (a b)^{-1} K$$

With the values from Table~\ref{t.model}, we find a brightness temperature for
the core component $T_B \ge 6.8 \times 10^9 K$. If we make the reasonable
assumption that the size of the emitting region is actually smaller (e.g.\ 1/3
of the deconvolved size), we get $T_B \ga 6 \times 10^{10} K$. However, even
under this assumption, the result requires a high but not extreme value of the
Doppler factor at the base of the radio jet. In fact, this is not surprising,
since the brightness temperature depends only on the physical length of the
maximum baseline and on the observed brightness. Therefore, observations with
similar array at a frequency near the spectral peak can yield higher $T_B$.

\begin{figure}
\resizebox{\hsize}{!}{\includegraphics{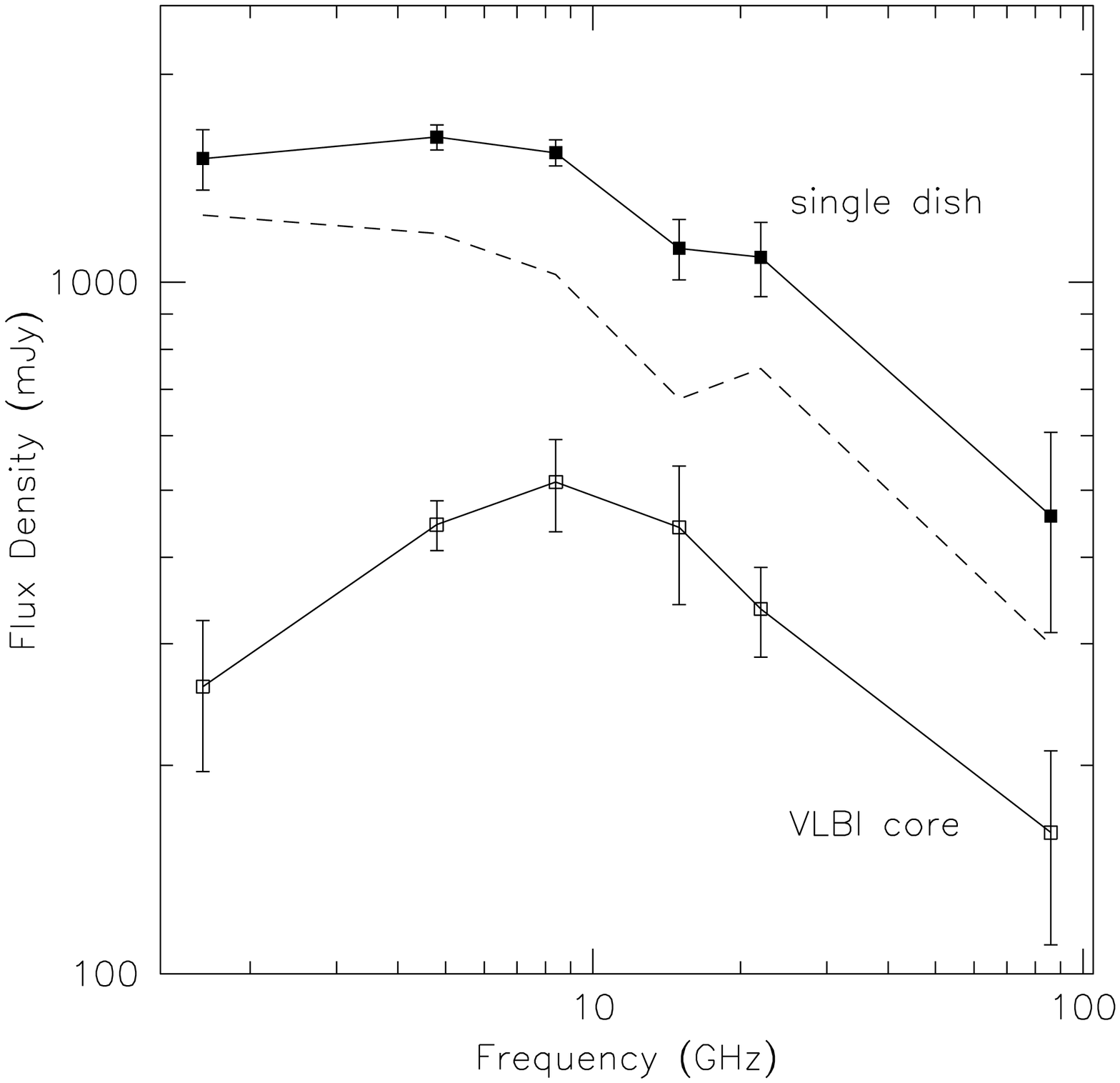}}
\caption{Spectra of Mrk~501. Filled squares represent average single dish
measurements \citep{ven01,all99,owe78,joy76,ste88}; empty squares show the VLBI
core flux: data between 1.6 and 22 GHz are mean values taken from
\citetalias{gir04a}; the datum at 86 GHz is from the present work. Solid lines
connect the points simply to guide the eye; the dashed line is the difference
between the two. Error bars show the standard deviation of averages but for the
86 GHz VLBI datum (instrumental calibration uncertainty).\label{f.spettro}}
\end{figure}

We show in Fig.~\ref{f.spettro} a spectral plot for both the (average) total
flux density on kiloparsec scale as measured by single dish telescopes, and for
the VLBI core, including our new data point at 86\,GHz. The spectrum of the
VLBI core between 1.6 and 22\,GHz has been presented in \citetalias{gir04a}:
the core has a turnover at about 8 GHz, and then the flux density falls as a
power law of index $\alpha = 0.5$. Our new data point follows the optically
thin part of the core spectrum. Note that as 86\,GHz flux density we do not
just adopt the flux measure of the unresolved core in our image but we also
include the jet like structure and the diffuse component discussed before; this
is because observations at lower frequency do not have the angular resolution
to distinguish the different components.  This implies that the turnover
frequency at $\sim 8$\,GHz is related to the whole structure and not to the
86\,GHz core, whose self-absorption peak is probably located at higher
frequency.

We also plot in Fig.~\ref{f.spettro} (dashed line) the difference between the
total single dish flux density and the VLBI core one. Apart from some
fluctuations (due to variability and non simultaneous data), the extended
emission from the kiloparsec scale jet and lobes region must substantially
contribute to the flux density even at high frequency. Such extended emission
has a rather flat spectrum, with index $\alpha = 0.3$ between 1.4 and 86\,GHz,
and fluctuations; this odd behavior is likely brought about as a consequence of
variability of the core.

The detection of a possible jet knot at $(r,\, \theta) = (0.72\, \mathrm{mas},
\, 172^\circ)$ is in agreement with images at lower frequency. Comparing the
GMVA data (Table~\ref{t.model}) with modelfits at 15 and 22\,GHz by
\citet{edw02} and \citetalias{gir04a}, the tentative jet knot can be identified
with the region labelled as C4, which is found at the same distance and
position angle.  \citet{edw02} found this region to be apparently stationary
between 1995 and 1999, and the positional coincidence lends support to this
interpretation. At this resolution, the jet has therefore a still different
orientation with respect to that seen at lower resolutions (see
\citetalias{gir04a} and Sect.\ \ref{s.jet}): the jet PA is $\sim 100^\circ$ at
$2<r<20$ mas and $\sim45^\circ$ at $r>20$ mas.

\subsection{Polarization}


\begin{figure*}
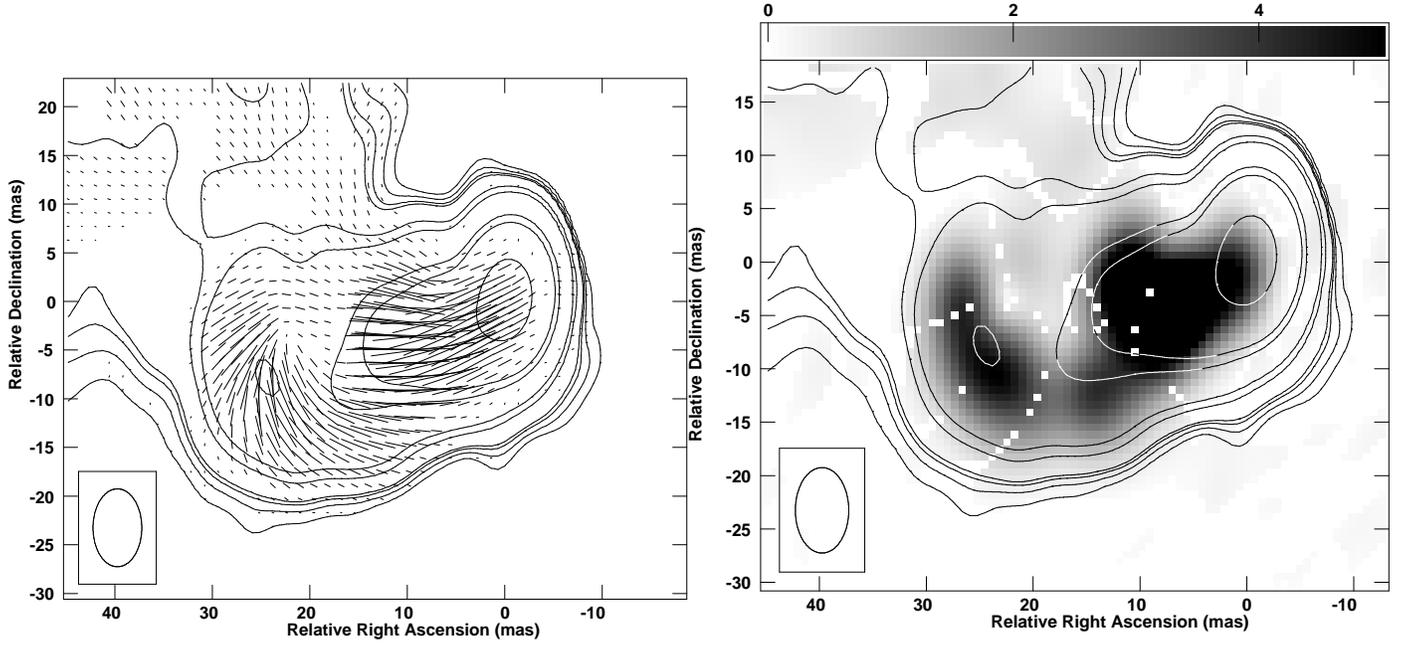

\resizebox{0.5\hsize}{!}{\includegraphics{9784f9a.eps}} \resizebox{0.5\hsize}{!}{\includegraphics{9784f9b.eps}}
\caption{Polarization images of the inner jet in Mrk 501, overlayed to total
intensity contours at 1.4 GHz, traced at (0.2, 0.5, 0.8, 1, 3, 5, 30, 50, 300)
mJy beam$^{-1}$. Left: EVPA, where 1 mas = 1 mJy beam$^{-1}$; right: polarized
intensity in grey tones between 0 and 5 mJy beam$^{-1}$.\label{f.polin}}
\end{figure*}

\begin{figure}
\resizebox{\hsize}{!}{\includegraphics{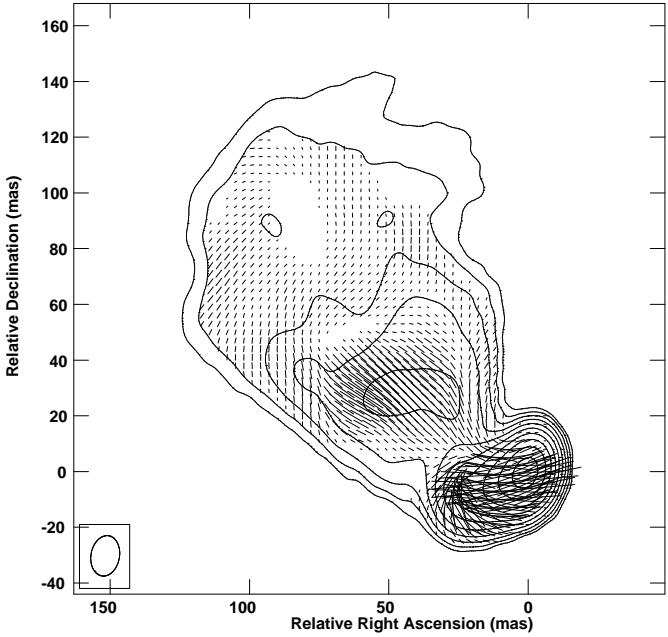}}
\caption{Polarization vectors overlayed to total intensity contours in the extended jet. Contours are traced at $(-1, 1, 2, 4, 8) \times 4.0$ mJy beam$^{-1}$, polarization vectors are plotted with a scale of 1.5 mas mJy$^{-1}$, where unpolarized flux $> 2$ mJy and polarized flux $> 0.3$ mJy.
\label{f.polout}}
\end{figure}

In polarized intensity, previous VLBI observations of Mrk~501 have revealed
flux densities of a few milliJansky, i.e., a few percent of the total intensity
\citep{pus05}. Our new HSA observations confirm the presence of a significant
fraction of polarized flux and reveal interesting details (see
Figs.~\ref{f.polin} and \ref{f.polout}). The total flux density in polarization
images is almost 100\,mJy (98.7\,mJy), with a peak of 18 mJy beam$^{-1}$. We
plot electric vector polarization angle sticks, assuming a Faraday rotation
measure of 0 rad m$^{-2}$.

In the inner 10 mas (Fig.~\ref{f.polin}), we have a large cone of polarized
flux, with polarization vectors aligned with the jet direction. Further
downstream, the polarized flux lies predominantly toward the southern edge of
the jet, in a structure similar to the `spine--sheath' detected in polarization
by \citet{pus05} and in total intensity by \citetalias{gir04a}. At the large
bending at 20 mas, we then find a knot of polarization and a rotation in the
EVPA position angle. This suggests a strong interaction between the radio
plasma and an external structure. This interaction could be the reason of the
change in the source direction, which is further amplified by geometrical
effects.

After the large bend (see Fig.~\ref{f.polout}), the polarization angle becomes
well aligned with the jet again at $\sim 100$ mas from the core. At larger
distances, the polarized signal becomes weak, with a modest preference for a
distribution on the south-east side and an orientation orthogonal to the jet
direction. Finally, the data from the VLA only (Fig.~\ref{f.vla}) confirm an
orientation of the electric vector parallel to the jet direction in the
extended jet, turning slightly clockwise at $r \ga 1\arcsec$.

\section{DISCUSSION}
\label{sec:discussion}

In \S\ref{sec:results}, we have presented our main new results about the core
and jet of Mrk~501. We now discuss their relevance for our understanding of the
physics of this source and of AGNs and jets in general.

\subsection{The inner core: radio core spectrum and GMVA structure}

The nuclear region of Mrk~501 consists of (1) an unresolved component: the
radio `core', point-like at our resolution (deconvolved size smaller than $\sim
30 \times 20 \, \mu$as or $0.020 \times 0.014$ pc or $200 \times 140 R_S$), and
(2) a faint resolved jet-like structure with a large opening angle, similar
(taking into account the significant difference in flux density and linear
resolution) to the inner structure of M87 \citep{ly07}.

The unresolved component with a total flux density of about 45 mJy, is
characterized by a relatively low $T_B$ in contrast with the higher $T_B$ of
M87. However, we note that because of the different distance from us of these
two sources, our `core' includes the whole M87 structure visible in images at
86 GHz \citep{kri06b}. Therefore, the $T_B$ of Mrk~501 is the average $T_B$ of
a resolved structure where the jet velocity could be only mildly relativistic,
as suggested by the detection of a counter-jet in the more misaligned M87
\citep{ly07,kov07}.

The diffuse low brightness emission with a total flux density of about 65 mJy
is interpreted as the continuation of the inner jet with spine--shear layer
structure. Better sensitivity images are necessary to properly map this low
brightness feature.

The lack of a dominant unresolved component is in agreement with the spectrum
shown in Fig.~\ref{f.spettro}, where we can assume that the observational data
refers to both regions (1) and (2) discussed above.  Of course, uncertainties
related to the variability are always present, but because of the regularity of
the spectrum we can use it to estimate the extension of the radio emitting
region, by simply inverting the following formula \citep{mar87}:

$$B = 3.2 \times 10^{-5} \, \theta^4 \, \nu^5_m \, S^{-2}_m \, \delta
\, (1+z)^{-1} $$

We assume a local average magnetic field $B = 0.02$ G and a Doppler factor
$\delta = 10$ \citep[as discussed by][]{tav01}.  The low ($\sim 8$ GHz)
self-absorption frequency requires that the emitting region has a size of the
order of 0.1 mas, in agreement with Fig.~\ref{f.gmva}, and that a point-like
source, if present, is not dominant.  Note that the angular size is
proportional to the magnetic field to the 1/4th power, therefore a relatively
small (even a factor 10) increase of the value of the magnetic field does not
affect these conclusions.

In our GMVA image, we also find a remarkable feature at $(r,\, \theta) =
(0.72\, \mathrm{mas}, \, 172^\circ)$. In fact, lower resolution VLBI
observations have shown so far that the jet of Mrk~501 does not have compact
jet knots on a few parsecs scale \citepalias{gir04a}.  If the new component is
indeed a jet knot, it will be important to re-observe the source and test if it
has a proper motion that can be followed. Alternatively, as suggested by the
positional coincidence with a feature in previous images, it could be a
standing shock at the position of a change in jet direction, due to unknown
reasons. Since this is not far from the first bend in the jet, it may be the
result of a standing shock from the disturbance causing the bend.

Moreover, beside this compact feature, a significant amount of flux density in
the GMVA data remains difficult to image and/or modelfit. This is mainly due to
the scarce coverage of the $(u,v)-$plane arising from some telescope failures.

\subsection{Jet structure and polarization} \label{s.structure}

Limb brightening in the jet of Mrk~501 seems to be present on scales as small
as 0.1 mas, but also after the two main bends at $\sim 2$ and $\sim 20$ mas,
where the jet has significantly expanded transversely. Under a given viewing
angle, different Doppler factors can arise from different velocities;
therefore, a common explanation for limb brightening in jets lies in the
existence of a velocity structure transverse to the jet, with an inner spine
and an outer shear.

The issue of the presence of transverse velocity structures is widely debated
at the moment, in the light of both analytical models and of numeric
simulations \citep{har07,miz07,per07,gop07,ghi05}. \citet{chi00} also suggested
the presence of transverse velocity structures to reconcile observational
results with the usually adopted AGN unification model. Direct evidence of limb
brightening has recently been confirmed, e.g.\ on the sub-parsec scales in M87
at 43 GHz \citep{ly07} and on a few parsecs scale in 1144+35 \citep{gio07}.
Mrk~501 is unique in the fact that the limb brightened structure is visible on
scales over three orders of magnitude, and in sections of the jet that are
differently oriented on the plane of the sky. In particular, the limb
brightening in the mm-VLBI image indicates that a spine/shear structure could
be present from the very inner part of the jet, close to the region where the
higher frequency optical emission is produced. Our image provides support for
the transverse dual velocity structure hypothesized by \citet{chi00} on the
basis of the correlation between the radio and the optical core luminosity in
radio galaxies and BL Lacs.

We note that despite the large changes in jet PA, the jet direction appears to
be constant in time; i.e., we do not have evidence of a change in the direction
of jet launch, of precession, or of variable activity. Only the sharp
brightness decrease at $\sim100$ mas from the core could be due to a
discontinuity in the jet activity but the evidence is marginal at
most. \citet{con95} tried to interpret this strucutre in terms of a geometrical
saturating helix model. However, the sensitive HSA image reveals that the
emission comes from a conical surface out to more that 700 mas, much further
out than the transition to cylindrical surface required in the saturating helix
model; moreover, in our image as well as in other sensitive low frequency
images \citep{gir04a,pus05}, the presence of the counter-jet feature described
by \citet{con95} and predicted by their model is excluded at high confidence
levels. Clearly, helicity remains an interesting explanation for the large
position angle changes in the jet ridge line \citep[see, e.g., the binary black
hole system described by][]{vil99}.  However, the necessary parameters can not
be constrained by the available images and the physical reality has to be more
complex than the geometrical model. \citet{lai06} have also ruled out a helical
magnetic field configuration in the nearby, low-luminosity radio galaxy 3C
296. The role played by interactions with the surrounding medium is surely non
negligible.

The polarization structure (Fig.~\ref{f.polin}) is also suggestive of
spine-sheath structures. The most prominent feature in our images is the
polarized cone departing from the inner region in E-W direction. We note that
because of the lower resolution, this cone extends over the full region
resolved in the images at higher frequency \citep{pus05}.  In our images,
polarization vectors are predominantly aligned with the jet axis, in contrast
with the dominant polarization in the external shear perpendicular to the jet
found by \citet{pus05}. Because of different frequency and resolution, a
comparison of the datasets is not obvious; one can assume that the difference
in the polarization vector orientation is mainly due to Faraday Rotation or
that at 1.4 GHz the dominant polarized flux is from the jet inner spine, and
this polarized flux has vectors oriented along the jet direction.

In our image, vectors are aligned with the jet direction also in the extended
jet structure in NS direction (after the large bending at about 50 mas from the
core). In this region, the central spine is again polarized with vectors
aligned with the jet axis and in the external shear a marginal evidence of
polarized flux oriented perpendicular to the jet is present.  Only in the
bending region vectors have a peculiar circular trend but here projection
effects can be dominant; moreover, in this region a strong interaction with the
ISM is present and it is the most likely region of the change of the jet PA.

We note that also vectors in the VLA map show a similar orientation if we
consider that the peak in the VLA polarization map is not at the core position
but is located in the main pc scale region at about or more 100 mas from the
core.  We conclude that at relatively low resolution and at 1.4 GHz the
dominant magnetic field structure is perpendicular to the jet axis (E vectors
aligned with the jet direction).

\subsection{Jet velocity and orientation}

\begin{figure*}
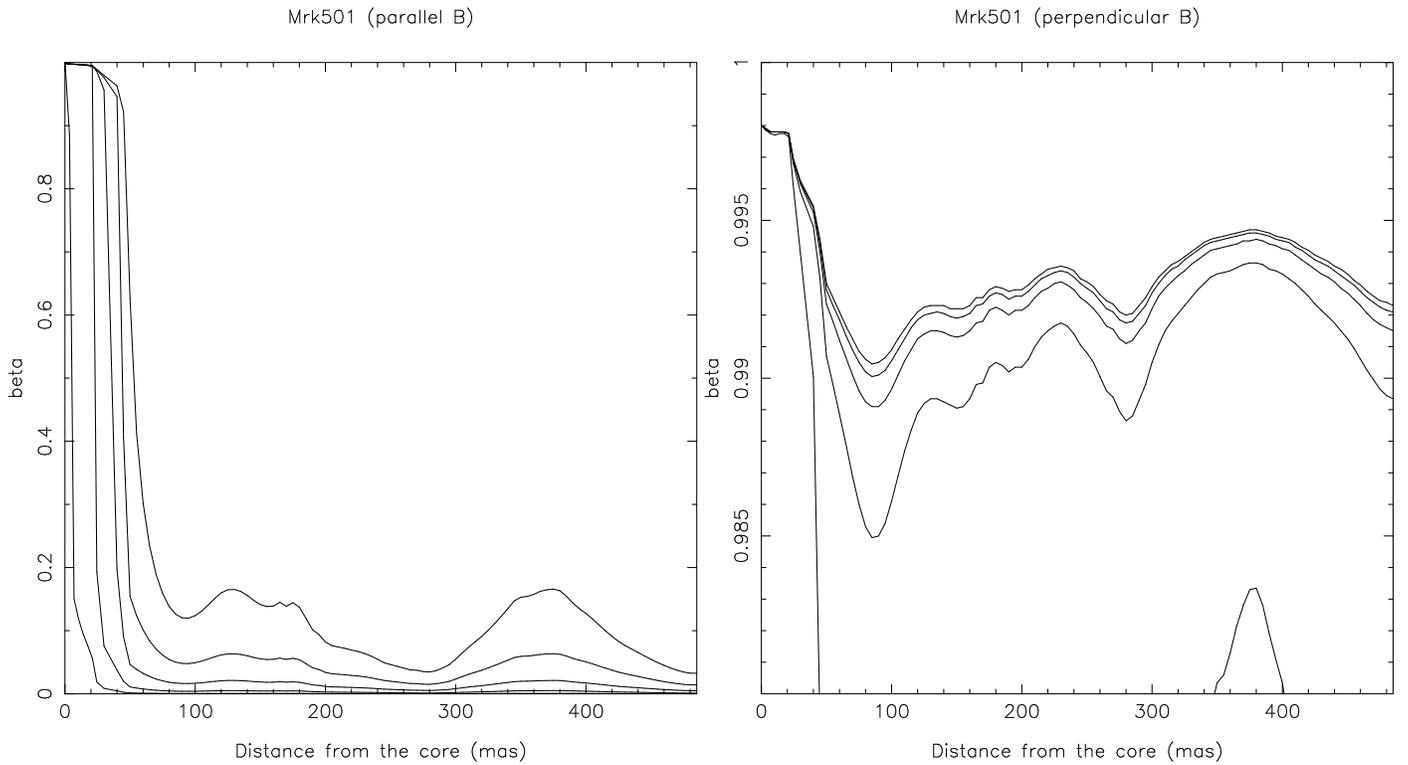

\resizebox{0.5\hsize}{!}{\includegraphics{9784f11a.eps}} \resizebox{0.5\hsize}{!}{\includegraphics{9784f11b.eps}}
\caption{Estimated jet velocity in the case of parallel (left) and
perpendicular (right) magnetic field. The initial velocity is
$\beta=0.998$. Viewing angles of
$\theta=5^\circ,10^\circ,15^\circ,20^\circ,25^\circ$ are shown, with smaller
angles at the bottom. Note the different scale on the $y-$axis, due to the
faster decrease of the jet speed in the case of parallel magnetic
field.\label{f.adiabatic}}
\end{figure*}

Our results show that the jet in Mrk~501 is characterized by different
properties on the various scales from a few hundreds to several millions
Schwarzschild radii. The jet orientation and velocity, and the ratio between
spine and shear contributions must significantly change over these scales. It
is therefore impossible to describe it with constant parameters.

Since a counter-jet is not detected, however, the jet has to be in a
relativistic regime even in its faintest and most extended region seen in the A
configuration VLA observation. The arcsecond scale structure is symmetric, so
we argue that the transition to non relativistic velocity has to occur
somewhere between projected distances of 1 to 10 kpc.

The jet opening angle remains constant ($\phi_\mathrm{j} \sim 40^\circ$) for
several 100 parsecs and any possible re-collimation can take place only in the
region where the jet becomes confused with the noise. Under our estimated
viewing angle for this part of the jet ($\theta \sim 15-20 ^\circ$), we derive
an intrinsic opening angle of $\phi^\prime_\mathrm{j} \sim 10-15^\circ$. Jets
with larger opening angles are found to have lower apparent speeds and Doppler
factors in analytical models \citep{gop07}. This may explain why Mrk~501 and
other TeV blazars do not tend to show strong superluminal motions
\citep{pin04}. However, it is to be noted that the HSA jet of Mrk 501 is
several beams wide and does not show evidence of any features such as the knots
considered in the analytical modeling of \citet{gop07}.

The results from the fits of an adiabatic model to the observed jet radius and
peak brightness presented in Sect.~\ref{s.jet} can be used to constrain the
bulk velocity, the orientation, and the magnetic field orientation in the
various parts of the jet. As we have discussed in the section about the
polarization properties (\ref{s.structure}), we have evidence that the magnetic
field is predominantly orthogonal to the jet axis.

First, we recall that from \citet{gir04a} it was shown that low initial Lorentz
factors ($\Gamma < 5$) are ruled out, regardless of the magnetic field
orientation, since they disagree with both the observed limb-brightened
structure and the jet/counter-jet ratio; moreover, they require a jet
deceleration between the $\gamma-$ray region and radio jet region that is too
strong.

In Fig.~\ref{f.adiabatic} we use the new HSA data to show the estimated jet
velocity in the case of parallel and perpendicular magnetic field, assuming an
initial $\beta=0.998c$ and an injection spectral index $\delta = 2$ (in
accordance with \citetalias{gir04a}). In each plot, we draw five lines,
corresponding to angles to the line of sight of 5$^{\circ}$, 10$^{\circ}$, ...,
25$^{\circ}$ (i.e., in the range of values allowed by the jet sidedness and
core dominance).

With these data, we now also rule out models starting with $\Gamma = 15$ with
magnetic field parallel to the jet axis, since our lower limit on the
jet/counter-jet brightness ratio near the core ($R>4000$), and at large
distance ($R>140$ at 120 mas) is inconsistent with the velocity decrease
predicted by a parallel magnetic field adiabatic model at $\sim 100$ mas. This
conclusion is also in agreement with the results derived on the polarization
properties, i.e., a magnetic field in the jet spine orthogonal to the jet
direction (EVPA parallel to the jet axis).

The fit with perpendicular magnetic field and an initial Lorentz factor $\Gamma
= 15$ is in general consistent with the other observational constraints. Only
in the case of the smallest viewing angle (i.e.\ $\theta = 5^\circ$) the jet
velocity falls off rapidly after the main jet bend; in the extended part of the
jet, narrow viewing angles are therefore not acceptable. However, it is
possible that the jet is more closely aligned in its inner part and then it
becomes oriented at a larger $\theta$ after the turn. For all the other viewing
angles, the fit velocities behave in rather similar ways. After an initial
decrease ($\beta = 0.985 - 0.991$, $\Gamma = 5.8 - 7.5$), the jet velocity
remains relativistic, with small oscillation. This is also in agreement with
the fact that the jet is still one-sided even on scales of a few kiloparsecs.

Finally, we note that the jet FWHM is probably underestimated at large $r$. For
this reason, we have also tried a fit with an input FWHM twice the measured one
in the outer jet. The results are qualitatively similar to the previous ones: a
parallel magnetic field is not acceptable and the perpendicular field implies
an initial deceleration and a more or less constant velocity further
out. Although the jet velocity is slightly smaller ($\Gamma = 4 - 6$), it still
remains in the relativistic regime.

\section{CONCLUSIONS}
\label{sec:conclusions}

We have successfully explored new regions in the remarkable jet of
Mrk~501. Thanks to the great sensitivity of the HSA, we reveal that the VLBI
jet is one-sided (and therefore in the relativistic regime) out to at least 500
parsecs from the core. The polarization vectors are clearly aligned with the
jet spine, suggesting that the magnetic field is orthogonal to the jet main
axis. This is also in agreement with the results of the adiabatic fit to the
jet brightness and width as a function of distance from the core. Limb
brightening -- already detected on intermediate scales by VSOP observations
\citepalias{gir04a} -- is now visible on HSA transverse profiles at $\sim 60$
mas from the core, and is likely present even on the sub-parsec scales imaged
by the GMVA.

Despite its presumed weakness, Mrk~501 has in fact been clearly detected by the
GMVA on sub-milliarcsecond scales. This result is encouraging given the
performance of the existing mm-VLBI array and suggests that not only the
brightest AGN can be studied on the smallest scales. Present and future
upgrades of the array (e.g., the installation of new receivers at Plateau de
Bure, inclusion of new or existing telescopes such as the 40\,m at Yebes,
Spain) are expected to make the instrument even more sensitive and
reliable. The brightness temperature of the most compact component is about $6
\times 10^{10}$\,K. This region has a linear size of $0.020 \times 0.014$\,pc
or, in terms of gravitational radii, $200 \times 140 R_S$. Significant emission
is also revealed in the short baselines between the most sensitive telescopes,
and awaits proper imaging with increased fidelity.

\begin{acknowledgements} 
We thank Dr.\ Luigina Feretti for useful discussions. We also thank the
personnel of the observatories participating in the Global mm-VLBI array and
particularly T.\ Krichbaum for his advice during the data reduction.  MAPT
research is funded through a Ramon y Cajal Fellowship from the Spanish
Ministery of Education.  The National Radio Astronomy Observatory is operated
by Associated Universities, Inc., under cooperative agreement with the National
Science Foundation. This research has made use of NASA's Astrophysics Data
System Bibliographic Services and of NASA/IPAC Extragalactic Database (NED)
which is operated by the Jet Propulsion Laboratory, California Institute of
Technology, under contract with the National Aeronautics and Space
Administration.


\end{acknowledgements}

\end{document}